\documentstyle [12pt]{article}
\newcommand{\bce}{\begin{center}}
\newcommand{\ece}{\end{center}}
\newcommand{\beq}{\begin{equation}}
\newcommand{\eeq}{\end{equation}}
\newcommand{\bef}{\begin{figure}}
\newcommand{\eef}{\end{figure}}
\newcommand{\bea}{\vspace{0.25cm}\begin{eqnarray}}
\newcommand{\eea}{\end{eqnarray}}

\newcommand{\mpi}{m_\pi}

\newcommand{\ba}{\begin{array}}
\newcommand{\ea}{\end{array}}

\newcommand{\ie}{{\sl i.e.~}}

\newcommand{\etc}{{\sl etc.~}}

\newcommand{\ave}[1]{\langle {#1} \rangle}
\newcommand{\singlespace}{
    \renewcommand{\baselinestretch}{1}\large\normalsize}
\newcommand{\doublespace}{
    \renewcommand{\baselinestretch}{1.6}\large\normalsize}

\newcounter{subeq}
\begin{document}
\pagestyle{empty}
\singlespace
\vspace{1.0in}
\begin{flushright}
June, 1993 \\
\end{flushright}
\vspace{1.0in}
\bce
{\large{\bf SELFCONSISTENT DESCRIPTION OF A THERMAL PION GAS}}
\vskip 1.0cm
R. RAPP and J. WAMBACH
\footnote{also: Department of Physics,
University of Illinois at Urbana-Champaign,
1100 West Green St.,\\
\indent
\hspace{0.42in} Urbana, IL 61081, USA\\
\indent ~~~and:
Institut f\"ur theoretische Kernphysik, Universit\"at Bonn,
D-5300 Bonn, F.R.G.}

\vspace{.35in}
{\it Institut f\"ur Kernphysik\\
Forschungszentrum J\"ulich\\
D-5170 J\"ulich, F.R.G }
\ece
\vspace{.65in}
\begin{abstract}
\noindent
We examine a hot pion gas by including medium modifications of
the two-body scattering amplitude as well as mean-field effects
selfconsistently. In contrast to earlier calculations, the in-medium
T-matrix is rather close to the free one while the mean-field potential
agrees well with lowest-order estimates. We also discuss the validity of
the quasiparticle approximation. It is found that it is reliable for
temperatures up to $\sim 150$ MeV. Above this temperature off-shell
effects in the pion selfenergy become important, especially
if the pions are strongly out of chemical equilibrium.
 \end{abstract}
\vspace{.75in}
\begin{flushright}
\singlespace
PACS Indices: 13.75.Lb\\
13.85.--t\\
14.40.Aq\\
\end{flushright}
\newpage
\pagestyle{plain}
\baselineskip 16pt
\vskip 48pt

\newpage
\doublespace

\section{Introduction}

 In ultrarelativistic heavy ion collisions several hundred particles
are produced in the final state. In the midrapidity region at CERN
energies of 200 GeV/A most of them are pions (with a pion-to-baryon ratio
of $\sim 6:1$). The lifetime of the
pionic fireball is a few fm/c and the freeze-out volume is typically
$10^3$ fm$^3$. The theoretical understanding of such a thermal 'pion gas'
is currently of great interest in connection with possible signals from
the quark-gluon plasma (QGP) phase transition \cite{StYo}.
On the other hand a hot and dense
Bose gas is also interesting from a many-body point of view since one
may expect interesting correlations associated with the statistics.

Several aspects of the thermal pion gas have been discussed previously,
including mean-field effects \cite{Shur} and medium modifications of the
$\pi\pi$ cross section \cite{BBDS}. It is the purpose of the present paper
to give a unified
description of these processes by using a reliable model for the
vacuum $\pi\pi$ interaction and requiring selfconstistency at the
two-body level. This leads to a Brueckner scheme, familiar
from the microscopic many-body theory of nuclear matter. Throughout
the discussion we shall assume that the gas is in thermal
but not necessarily in chemical equilibrium ($\mu_\pi\neq 0$). The latter
seems to be required from fits to pion $p_T$-spectra, which yield
$\mu_\pi\sim\mpi$ \cite{KaRu}. The assumption of thermal equilibrium,
on the other hand, seems reasonable from the following simple estimate:
at 200 GeV/A the freeze-out density $n_\pi$ of pions produced at
midrapidity is $\sim 0.3fm^{-3}$. With $N_\pi\approx 400$ this yields
a freeze-out radius $R_f\sim 7$ fm. Taking an average $\pi\pi$ cross
section $\ave{\sigma}=15$ mb gives a mean free path
$\lambda=1/n_\pi\ave{\sigma}\sim 2.2$ fm and hence the mean number of
collisions is $\sim 3$. Thermalization is further corroborated in a
scenario where initially a quark-gluon plasma (${\it QGP}$) is formed.
Here QCD string breaking models as well as parton cascade simulations
also yield thermal equilibration which should survive during the hadronization.

\section{The Vacuum $\pi\pi$ Interaction}

The starting point for our description of the hot pion gas is the
vacuum $\pi\pi$ interaction model of ref.~\cite{LDHS}
which is based on meson exchange. Here the basic meson-meson interaction
is constructed from an effective meson Lagrangian  with
phenomenological form factors at the vertices.
{}From these vertices two-body pseudopotentials
are constructed including the most important {\it s}- and {\it t}-channel
meson exchanges. Among these the {\it t}-channel $\rho$ exchange between
$\pi\pi$ states and the {\it s}-channel $\rho$ pole term will be most
important for our discussion. We employ the Blankenbecler-Sugar (BbS)
reduction \cite{BbS}  of the 4-dimensional Bethe-Salpeter equation
maintaining covariance \cite{PeHS}.
The $\pi\pi$ $T$-matrix for given angular momentum $J$ and
isospin $I$ is obtained as
\begin{eqnarray}
T_{\pi\pi}^{JI}(Z,q_1,q_2) & = & V_{\pi\pi}^{JI}(Z,q_1,q_2)+
     \nonumber\\
 & &
+ \int_0^{\infty}dk \ k^2 \ 4{\omega}_k^2 \ V_{\pi\pi}^{JI}(Z,q_1,k)
 \   G_{\pi\pi}^0(Z,k) \ T_{\pi\pi}^{JI}(Z,k,q_2)  \ ,
\label{eq:Tmat}
\end{eqnarray}
where $k=|\vec k|$ \etc ; $Z$ is the CMS energy
and $G^0_{\pi\pi}(Z,k)$ the vacuum two-pion propagator in the CMS
frame with pions  of momenta $\vec k$ and $-\vec k$ (the Lohse et al.
model also contains coupling to the $K\bar K$ channel which has been
omitted for brevity in eq.~(\ref{eq:Tmat}) but which is included in
the calculation). In the BbS form the two-pion propagator is given by
\beq
G_{\pi\pi}^0(Z,k)=\frac{1}{\omega_k} \  \frac{1}{Z^2-4\omega_k^2+i\eta}
\eeq
with ${\omega_k}^2=k^2+\mpi^2$.
This model gives a good description of the phase shifts and inelasticities
up to $\sim 1.5$ GeV which is more than sufficient for our purposes.

\section{Selfconsistency}

The most obvious medium modification of the $\pi\pi$ scattering in the
gas surrounding is a change in momentum weight of the intermediate
two-pion propagator, first studied in ref.~\cite{BBDS}.
Identifying the CMS frame with the thermal reference frame one has
\beq
G_{\pi\pi}(Z, k;\mu_\pi,T)=
\frac{1}{\omega_k} \ \frac{(1+2f_k(\mu_\pi,T))}
{Z^2-4\omega_k^2} \ ,
\eeq
where
$f_ k=(\exp \lbrack (\omega_k-\mu_{\pi})/T \rbrack -1)^{-1}$
is the thermal Bose factor and $\mu_\pi$  the chemical potential. The
identification of the CMS frame with the thermal frame simplifies our
calculations considerably (allowing for a relative velocity between
the two frames we find that it effectively acts like a change in
$\mu_\pi$ which is a parameter for us anyway).
At fixed $T$ the pion chemical potential fixes the density via
\beq
n_\pi(\mu_\pi,T)=g_\pi \int \frac{d^3q}{(2\pi)^3}  \ f_q(\mu_\pi,T) \ ,
\eeq
where $g_\pi=3$ is the isospin degeneracy factor.
We use a temperature range of 100-200 MeV.
A temperature of 100 MeV is roughly
the lowest temperature from thermal fits to $p_T$-spectra at the AGS. On
the other hand 200 MeV should be an upper limit for purely hadronic models
since one expects the phase transition into a QGP around this value.
While thermal equilibrium seems a reasonable assumption it is not clear
whether chemical equilibration is reached during the evolution of the
pion gas. Indeed, fits of the CERN $p_T$-spectra could be improved with
$\mu_\pi\sim 125$ MeV \cite{KaRu}. Hence we discuss both cases $\mu_\pi=0$
and $\mu_\pi=125$ MeV. At the same temperature the latter gives a higher
pion density reaching a maximal density $n_\pi\sim 0.7$ fm$^{-3}$ at
200 MeV.

There is a second effect which needs to be considered \cite{Shur}.
The $\pi\pi$ interaction introduces a pion selfenergy $\Sigma_\pi$
('mean field') which changes the single-pion dispersion relation
\beq
\omega^2_k=m^2_\pi+ k^2+\Sigma_\pi(\omega_k,k;\mu_\pi,T)
\eeq
as a function of density and temperature.
In terms of the forward-scattering amplitude $M_{\pi\pi}$, $\Sigma_\pi$
is expressed as \cite{Shur,Sche} :
\begin{eqnarray}
\Sigma_{\pi}(\omega,k;\mu_{\pi},T)=\int \frac{d^3p}{(2\pi)^3}\frac{1}{2
\omega_p}  \ f_p(\mu_\pi,T) \ M_{\pi\pi}(k^{(4)},p^{(4)}) \ .
\label{eq:Self}
\end{eqnarray}
Relating the forward-scattering amplitude to the T-Matrix as
\beq
M_{\pi\pi}(E_{cms})=(2\pi)^3 \ E_{cms}^2 \ T_{\pi\pi}(E_{cms}) \ ,
\eeq
where $E_{cms}=\sqrt s =\lbrack (p+k)^\mu \ (p+k)_\mu \rbrack ^{1/2}$ is
the CMS energy of the two colliding pions, one can transform the
selfenergy expression into
\beq
\Sigma_\pi(\omega,k;\mu_\pi,T)=\frac{\pi}{k} \int\limits_0^\infty dp \
\frac{p}{\omega_p} \ f_p(\mu_\pi,T) \int
\limits_{E_{min}}^{E_{max}} dE_{cms} \ E_{cms}^3 \ T_{\pi\pi}(E_{cms}) \ .
\label{eq:Sig2}
\eeq
Here we have restricted ourselves to the on-shell T-Matrix
neglecting the dependence on the total momentum $\vec P=\vec k +\vec p$
of the pair. Thus the energy integration bounds are given as
\beq
E_{max/min}=(\omega^2+\omega_p^2+2\omega\omega_p-k^2-p^2\pm2kp)^{1/2} \ .
\eeq
For the forward-scattering T-Matrix we take the spin-isospin weighted sum
including partial waves up to $J=2$:
\beq
T_{\pi\pi}(E_{cms})=\frac{1}{4\pi} \sum_{I,J=0}^{2} \
\frac{(2I+1)}{3} \ (2J+1) \ T_{\pi\pi}^{JI}(E_{cms}) \  ,
\eeq
which saturates the cross section in the relevant energy range.

It should be noted that the rate $\Gamma_k=-2Im \Sigma_\pi
/(2(k^2+m_\pi^2)^{1/2})$ deduced
from eq.~(\ref{eq:Sig2}) is not entirely consistent with that obtained
from the collision term of the bosonic Boltzmann equation. A correct
account of the Bose correlations of the two interacting pions will
modify the occupancy factor in (\ref{eq:Sig2}) \cite{ChDa}.
It turns out, however, that this is a small effect \cite{Chan}.

It is now evident that the pion selfenergy and the T-Matrix should be
combined in a selfconsistent Brueckner scheme as indicated in Fig.~1. This
implies that the selfenergy (8) is to be calculated from
the in-medium
T-Matrix which, on the other hand, should be obtained from the in-medium
2$\pi$ propagator including the pion selfenergy:
\beq
G_{\pi\pi}(Z,k;\mu_\pi,T)=(1+2f_k(\mu_\pi,T)) \ \int
\frac{id\omega}{2\pi} \ D_\pi(\omega,k) \ D_\pi(Z-\omega,k) \ ,
\eeq
where
\beq
D_\pi(\omega,k)=\lbrack \omega^2-m_\pi^2-k^2-\Sigma_\pi(\omega,k;\mu_\pi,T)
\rbrack ^{-1} .
\eeq
Together with the in-medium scattering equation,
\beq
T_{\pi\pi}(\mu_\pi,T)=V_{\pi\pi}+V_{\pi\pi}
\ G_{\pi\pi}(\mu_\pi,T) \ T_{\pi\pi}(\mu_\pi,T) \ ,
\eeq
equations (8), (11) and (12) define the selfconsistency problem (see
also Fig.~1). It should be noticed that the pseudopotential $V_{\pi\pi}$
remains unchanged.

In the following we will discuss two different methods of calculating
$G_{\pi\pi}$.

\section{Quasiparticle Approximation}

The quasiparticle approximation (QPA) is valid if the lifetime is long
or more precisely if the quasiparticle energy
\beq
e_k\equiv (m_\pi^2+k^2+Re \Sigma_\pi(e_k,k))^{1/2}
\eeq
is much larger than its decay width $\Gamma_k$.
In this case the energy-dependence of $Re\Sigma_\pi$ is expanded to
first order around the 'quasiparticle pole' $e_k$ as
\beq
Re\Sigma_\pi(\omega,k)\approx Re\Sigma_\pi(e_k,k)+
\frac{\partial Re\Sigma_\pi(\omega,k)}{\partial\omega^2}|_{e_k}
 \ (\omega^2-e_k^2) \ .
\eeq
One can then perform the folding integral (11) analytically which gives
\beq
G_{\pi\pi}(Z,k;\mu_\pi,T)=\frac{1}{\bar \omega_k} \frac{z_k^2 \ (1+2
f_k(\mu_\pi,T))}{Z^2-4\bar \omega_k^2}
\eeq
with
\bea
z_k & \equiv & (1-\frac{\partial Re\Sigma_\pi(\omega,k)}
{\partial\omega^2}
|_{e_k})^{-1} \quad {\rm the \ pole strength} \ ,
\nonumber\\
\bar \omega_k^2 & \equiv & e_k^2 + i \ z_k \
Im\Sigma_\pi(e_k,k) \quad
{\rm quasipion \ dispersion \ relation} \ .
\eea
Together with this approximation eqs. (8),(11) and (13) are solved by iteration
starting from the free pion dispersion relation and the vacuum T-Matrix.
We keep the pion density $n_\pi$ fixed during the iteration by
readjusting $\mu_\pi$ in each step (the final $\mu_\pi$ differs from the
starting value $\mu_\pi^{(0)}$ by a small amount).
The selfconsistent results are shown in Fig.~2.
Already for chemical equilibrium ($\mu_\pi^{(0)}=0$) we find a
considerable reduction of the peak values in $Im T^{JI}$ as
compared to the vacuum case in both s- and p-wave.
This is in agreement with the lowest-order results from refs.~\cite{BBDS,
ACSW}.
The reduction is mainly caused by the Bose factors $(1+2f)$ leading to
a stronger weighting of the lower pion energies. For the same reason the
near threshold region shows an enhancement over the vacuum T-Matrix
especially in the s-wave. We define the pion 'optical potential' as
\cite{Shur}
\beq
V_\pi(k)\equiv \frac{\Sigma_\pi(e_k,k)}{2(k^2+\mpi^2)^{1/2}} \ ,
\eeq
which is shown in the lower part of Fig.~2.
In agreement with refs.~\cite{Shur,Sche}
we find attraction in $Re V_\pi$ for low momenta. The selfconsistent
potential, however, is significantly weaker than in lowest order.
The maximum in $Im V_\pi$ is
due to formation of the $\rho$-resonance, whereas the non-zero
values at $k=0$ arise from s-wave interaction with thermally moving
pions.

\section{Off-shell Integration of $G_{\pi\pi}$}
To check the reliability of the QPA we
perform the same calculations as described in the previous
section, but the quasiparticle two-pion propagator is now replaced by a
numerical integration of (11) accounting for the full off-shell
properties of $\Sigma_\pi(\omega,k)$. Using the symmetry relation
\beq
\Sigma_\pi(-\omega,k)=\Sigma_\pi(\omega,k) \ ,
\eeq
the in-medium propagator (11) can be written as
\beq
G_{\pi\pi}(Z,k;\mu_\pi,T)=(1+2f_k(\mu_\pi,T))
\int\limits_{Z/2}^{\infty} \frac{i d\omega}{\pi} \ D_\pi (\omega,k) \
D_\pi(|Z-\omega|,k) \ .
\eeq
Fig.~3 shows the selfconsistent results
for a chemically equilibrated pion gas ($\mu_\pi^{(0)}=0$). At
$T=125$ MeV the results coincide  with the QPA within a few percent.
For $T\rlap{\lower 0.5 ex \hbox{$\sim$}}
\raise 0.5 ex \hbox{$>$}150$ MeV the deviations become
larger: the T-Matrices in s- and p-wave are now enhanced compared to
the vacuum curve for most of the energy range (e.g. the peak value of
the $\rho$-resonance increases by $\approx 30\%$ at $T=200$ MeV).
This clearly must be an off-shell effect. From the lower part of Fig.~3
one can conclude that a
first-order expansion of $Re\Sigma_\pi$ around $e_k$ does not
describe the energy dependence correctly; in addition
$\Gamma_k/e_k\sim 0.6$ around $k=300MeV/c$, \ie the quasiparticle
lifetime becomes very short and renders the QPA invalid.
The potentials $V_\pi(k)$ are now very close to the lowest-order results.
As compared to the QPA a considerable amount of attraction is restored
in the low-momentum region of $ReV_\pi(k)$.
We also investigate a scenario with finite chemical potential
$\mu_\pi^{(0)}=125MeV$ as suggested by thermal fits
to the SPS $p_T$-spectra \cite{KaRu}. The s-wave $\pi$$\pi$
interaction now shows a strong accumulation of strength near threshold
(Fig.~4) which might lead to quasi-bound $\pi^+\pi^-$ pairs
as suggested in ref.~\cite{ACSW}. The latter possibility
deserves further study. In the p-wave
the changes of the $\rho$-resonance are appreciable: at highest
temperatures the width decreases considerably which might be detectable
via dilepton pairs coming from the midrapidity region.

\section{Summary}
Based on a selfconsistent Brueckner theory we have presented a numerical
analysis of a hot, interacting pion gas in thermal equilibrium, which may
be realized in future experiments at RHIC or LHC (at
$\sqrt{s}\ge 100 $GeV/A). The selfconsistent Brueckner scheme accounts
for statistical (Bose factors)  as well as dynamical (selfenergy)
modifications of the pion propagation in the gas. Using a realistic model
for the vacuum $\pi$$\pi$ interaction we have solved the non-linear
problem by iteration. We have compared results from the QPA
to those taking full account of the off-shell behavior in the pion
selfenergy. It is found that the QPA breaks down
for temperatures $T\ge 150MeV$ and finite chemical potentials, \ie
high pion densities. In the full calculation we find a slight
{\sl increase} of the in-medium  T-matrix --  in contrast to
refs.~\cite{BBDS,ACSW}, where only statistical modifications have been
taken into account. The main cause of this enhancement is attributed to
strong off-shell effects in the pion selfenergy.  The single-particle
potentials show considerable attraction for low pion momenta, although
quantitatively somewhat less than our lowest-order results.
As was shown recently by Koch and Bertsch \cite{KoBe}, such an attraction
is not able to explain the low-$p_T$ enhancement in the pion spectra of
current experiments. However, as suggested by numerical simulations of
the bosonic transport equation \cite{BDSW}, an increase of the in-medium
T-Matrix, as found in our calculations, is likely to ensure
thermalization of the pionic fireball and hence to produce an excess of
low-$p_T$ pions through the $(1+f)$ factors in the collision integral.
This issue and the impact of finite baryon density in the pion gas
will be addressed to a future publication.

\bce
{\bf Acknowledgement}
\ece
\vskip 1truecm

\noindent
We thank G. F. Bertsch, G. E. Brown, G. Chanfray, M. Prakash,
E. V. Shuryak and P. Schuck for useful discussions.
This work was supported in part by a grant from the National Science
Foundation, NSF-PHY-89-21025.
\vfill\eject

\vfill\eject

\bce
{\bf FIGURE CAPTIONS}
\ece
\vskip 2.0cm
{\parindent = 0pt
{{\bf Fig.~1} Set of selfconsistent equations for an interacting
pion gas:
\newline
upper part: Dyson equation for in-medium pion propagation
\newline
lower part: in-medium $\pi$$\pi$ T-Matrix equation.

\vskip 0.5 cm

{{\bf Fig.~2} Selfconsistent pion gas at $\mu_\pi^{(0)}=0$ in QPA:
\newline
upper part: imaginary part of the $\pi$$\pi$ T-Matrix in the JI=00- and
JI=11-channels for several temperatures (long-dashed lines: $T=125$ MeV,
short-dashed lines: $T=150$ MeV, dotted lines: $T=200$ MeV)
\newline
lower part: corresponding pion potentials (short-dashed lines:
$T=150$ MeV, dotted lines: $T=200$ MeV; the dashed-dotted and
dashed-double-dotted lines show the lowest-order results calculated with
the vacuum T-Matrix for $T=150$ MeV and $T=200$ MeV, respectively).
\vskip 0.5 cm

{{\bf Fig.~3} Selfconsistent pion gas at $\mu_\pi^{(0)}=0$ with full
off-shell integration of $G_{\pi\pi} (eq.~(20))$:
\newline
upper part: imaginary part of the $\pi$$\pi$ T-Matrix (see Fig.~2)
\newline
lower part: real part of the pion potentials (see Fig.~2) and real part
of the off-shell pion selfenergy at $T=200$ MeV.
\vskip 0.5 cm

{{\bf Fig.~4} Imaginary part of the selfconsistent $\pi$$\pi$ T-Matrix
with full off-shell integration of $G_{\pi\pi}$ for
$\mu_\pi^{(0)}=125$ MeV (long-dashed lines: $T=100$ MeV, short-dashed lines:
$T=150$ MeV, dotted lines: $T=175$ MeV, full lines: free space).

\end{document}